# A Molecular Rotor Driven by an Electric Field on Graphene

Wanxing Lin,[†, ‡] Xiaobo Li,[§] and Yan-Ling Zhao [*, ‡, ‖]

[†] School of Materials Science and Engineering, Guangdong Ocean University, Yangjiang 529500, P. R. China

[‡] Department of Physics, City University of Hong Kong, Hong Kong SAR 999077, P. R. China

[§] School of Physics and Optoelectronic Engineering, Hainan University, Haikou 570228, P. R. China

[‖] Advanced Energy Storage Technology Research Center, Shenzhen Polytechnic University, Shenzhen 518055, P. R. China

[*] E-mail: zhaoyanling@szpu.edu.cn.

**Abstract**

We propose a scheme for driving a dipolar molecular rotor to rotate continuously by applying an external electric field: the dipolar rotor is fixed on a graphene sheet via a metal atom to facilitate the free rotation; it is in the meantime subjected to an electric field oriented parallel to the graphene sheet. We use computational modeling with density functional theory and Newtonian mechanics, similar to molecular dynamics simulations, to obtain the torque, angular velocity, and rotation period of the rotor. Our results show that the dipolar rotor designed here can rotate with a period of 2.96 ps by an alternating rectangular electric field with a strength of 0.5 V/Å. However, a cosine wave alternating electric field depending on time cannot drive the dipolar rotor to rotate regularly. Therefore, a cosine wave electric field depending on the rotation angle is suggested, as it can not only drive the rotor but also produce additional power. Machine learning molecular dynamics (MLMD) simulations further confirm that the rotor remains thermodynamically stable under an electric field. This work reveals the rotation mechanism of a dipolar molecular rotor in a transverse electric field, and we hope this work can open a new path for designing more diverse molecular machines in experiments.

# 1. Introduction

In recent decades, research on molecular machines has been a popular topic due to their unique structural features and potential applications in nanoscience.[1-6] In fundamental research, the molecular rotor serves as a valuable model to simulate the structural and dynamical mechanisms of molecular machines.[1] Although vibration, rotation, and translation are the intrinsic dynamic properties of molecules, rotation is the most concerned in the study of molecular rotors. Previous research shows that molecular rotors can be driven by light,[1,2,7-9] chemical energy,[10] electric fields,[11,12] STM tips,[5,13-19] or by electron gain and loss.[20] Recently, the rotational dynamics of large solid-state molecular rotors have been observed,[21] the DNA rotor on a nanopore presents unidirectional rotation by applied voltage,[22] the DNA rotor acts as an overdamped electrical motor at high electric field amplitudes.[23] A repetitive rolling of molecular wheels has been achieved on a Cu(610) substrate using the intermittent electric field,[24] and a rotating electric field can be used to manipulate droplets.[25] Remarkably, 3D printing paves a new avenue for designing molecular rotors.[26]

Regarding the theoretical study of molecular rotors, recent proposals include rotational structures built on copper surfaces and graphene sheets, with systematic investigations into their rotational mechanisms. If some molecular rotors are rotated manually, they can drive neighboring rotors through repulsive forces.[27-29] These studies explain the rotational cooperation among multiple small machine units but neglect to discuss the input driving force. When it comes to electric field input, we are interested in the behavior and properties of dipolar molecular rotors driven by rotating electric fields,[12,30] oscillating electric fields,[31] and terahertz electric fields.[32] The electric-field-induced torque effects on nanoparticles have also been further described by atomistic-level simulations.[33] It is interesting to explore how a unidirectional and periodic rotation of a dipolar rotor behaves in an electric field.

Molecular rotors with high structural symmetry and no dipole moments, such as benzene rings, cannot be driven by a transverse electric field because the external field does not generate any torque on the rotors. To develop a theoretical model, we designed an exemplary dipolar molecular rotor that includes an electron-donating group and an electron-withdrawing group. As is well-known, electron-donating functional groups can carry positive charges in a compound, while electron-withdrawing groups carry negative charges. When both are incorporated into a molecule, a permanent dipole moment forms, and the molecule is expected to rotate in a transverse electric field. To quantitatively understand the primary rotational mechanism of molecular rotors, a torque analysis method has been proposed and employed in several earlier theoretical and computational studies.[8,20,27-29,34] The behavior of gears has been investigated using theoretical simulations within a rigid-body approximation based on a few collective variables.[35] The STM experiment reveals that the electric field plays a central role in driving azulene-based single molecules on Au(111).[36]

In this work, a rotor with a dipole moment mounted on a finite graphene sheet is studied. Various schemes for driving the rotor to rotate are proposed and compared. The rotation mechanism on the graphene sheet is investigated using density functional theory (DFT) combined with Newtonian mechanics in an approach similar to molecular dynamics. Additionally, the rotational behavior of the rotor in the electric field has also been examined using MLMD simulations. We hope this work improves our understanding of how similar dipolar molecular rotors behave in transverse electric fields and facilitates their realization in experiment.

2. Modeling and Methods

The entire system investigated here is a sandwich structure, as shown in Figure 1. After complete optimizations using the DFT method based on the quasi-Newton (or variable metric) algorithm in the absence of electric fields, all atoms of the finite graphene sheet are nearly coplanar, and those in the rotor are also coplanar except for the two H atoms in the $-NH_2$ group. We define the configuration in Figure 1a as the 0° state, and during the simulation, we rotate the rotor counterclockwise, through the Cr atom, around the z direction. The perpendicular distance from the Cr atom to the graphene sheet $d_1$ is 1.706 Å, and the distance from the Cr atom to the benzene ring's center $d_2$ is 1.584 Å. At the 0° state, the absolute dipole moments of the isolated rotor along the x, y, and z directions are calculated to be 6.623, 0.005, and 1.282 D, respectively. When a fixed electric field of 0.5 V/Å is applied along the −y direction, the rotor tends to rotate counterclockwise around the rotation axis, through the Cr atom, without tilting caused by the torque on the dipolar rotor. The system was built using the Device Studio code,[37] and the atomic coordinates are shown in the Supporting Information. This molecule was previously proposed in our earlier work,[38] while we systematically and thoroughly investigate the rotation mechanism under the external electric field in this work.

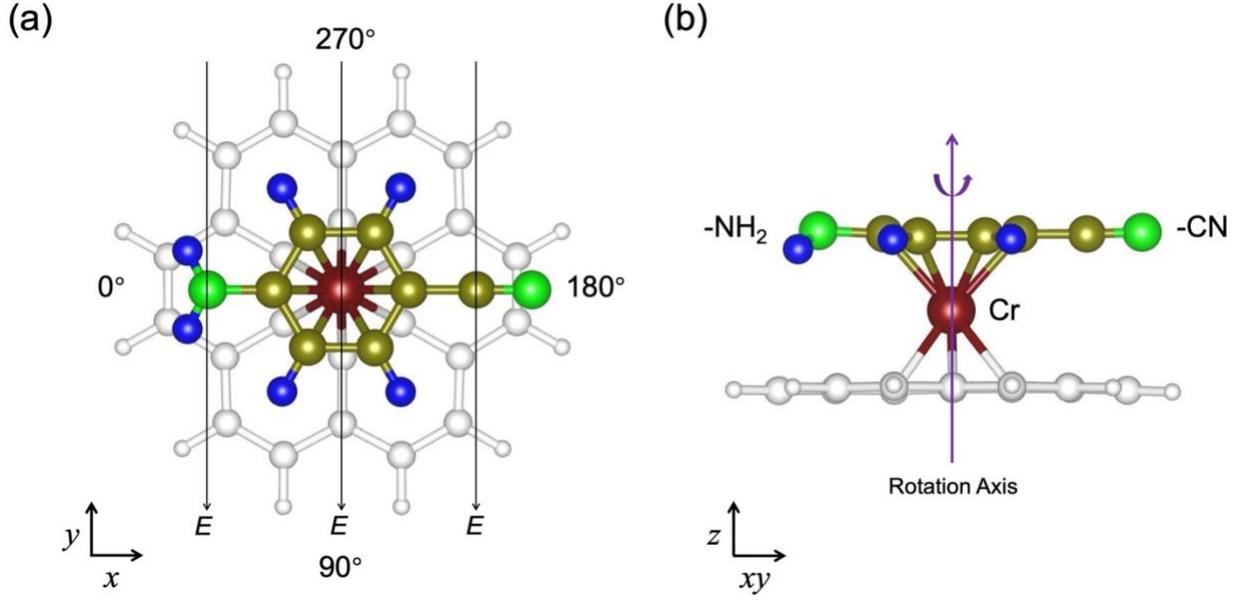

**Figure 1.** (a) The top view and (b) the side view of the sandwich structure, which includes a dipolar rotor, a Cr atom, and a finite graphene sheet with dangling bonds saturated by H atoms. The dark red sphere represents the Cr atom; the olive, green, and blue spheres represent the C, N, and H atoms in the rotor, respectively. The large white balls and small white balls represent the C and H atoms of the graphene sheet, respectively. The $E$ arrows indicate the electric field is applied along the $-y$ direction.

As the constant electric field drives the rotor (perpendicular to the rotation axis), the rotor's shape remains unchanged. Therefore, the simulation can be simplified as a rigid body rotation around the axis. We assume the torque is constant at each 1° step. Specifically, the angular acceleration $\alpha$ is constant within the interval from $[\theta, \theta + 1°]$. During rotation, the rotational inertia of the rotor is

$$I_{zz} = \sum_{i=1}^{n} m_i \times r_{z,i}^2, \quad (1)$$

where $i$ runs over all atoms of the dipolar rotor, and $r_z$ is the vertical distance from each atom to the rotation axis. The inertia of this rotor with respect to the rotation axis is: $I_{zz} = 1.046 \times 10^{-44}$ kg·m². The torque along the rotation axis $M_z$ can be calculated using the formula, $M_z = \sum_{i=1}^{n}(r_i \times F_i)_z$, where $r_i$ and $F_i$ indicate the position vector of atoms relative to the rotation axis and the forces on atoms, respectively. $F_i$ can be calculated by the self-consistent-field DFT calculation. Additionally, the angular acceleration $\alpha$ can be derived using the moment of momentum theorem

$$M_z = I_{zz} \times \dot{\omega} = I_{zz} \times \alpha, \quad (2)$$

where $\omega$ is the angular velocity. At any given moment, the angular velocity

$$\omega = \omega_0 + \frac{M_z}{I_{zz}} \cdot t \quad (3)$$

and the angle

$$\theta = \theta_0 + \omega_0 \cdot t + \frac{M_z}{2I_{zz}} \cdot t^2 \quad (4)$$

can be obtained accordingly.

The first principle calculations were performed using the Vienna *Ab initio* Simulation Package (VASP) with the projector augmented wave method,[39,40] and a cutoff energy of 400 eV was employed. The Perdew–Burke–Ernzerhof (PBE) functional was used as the exchange-correlation.[41] The vacuum distance between different layers is more than 16 Å. Before applying an electric field, the system was fully optimized using a quasi-Newton (or variable metric) algorithm until the net force on each ion was less than 0.01 eV/Å. In both $E = 0$ and $E = 0.5$ V/Å cases, the convergence criterion is $10^{-4}$ eV in the self-consistent-field calculations. The spin has been polarized, and van der Waals interactions were calculated using the Grimme method with a zero-damping function.[42] MLMD simulations were conducted with force fields generated on-the-fly through the Bayesian inference approach integrated within VASP 6.5.1.[43,44] The simulations were performed in the canonical (NVT) ensemble, where the graphene sheet and Cr atom were fixed, and the rotor was thermally activated at temperatures of 0, 50, 100, and 300 K using the Andersen thermostat.[45] Each simulation ran for 10,000 steps with a time step of 1 fs. Visualization of the simulation animations was done using OVITO (Open Visualization Tool),[46] which provided detailed insight into the rotor's structural evolution and dynamic behavior under an electric field.

Initially, the system was fully optimized without an electric field until reaching equilibrium, as shown in Figure 1. In the second step, we fixed the graphene sheet and relaxed the dipolar rotor and the Cr atom in a constant electric field of 0.5 V/Å applied along the −y direction. It was found that the rotor can rotate around the rotation axis through the Cr atom without tilting. In the third step, we manually rotated the rotor from 0° counterclockwise in 5° increments around the rotation axis while maintaining the electric field along the −y direction. We performed self-consistent-field calculations for each configuration, obtaining the total energy of the system and the forces on each atom. This step involves no structural optimization, so the rotation is rigid. Vibrations, translations, and flexibility were neglected because their effects are minimal.[27,29]

## 3. Results and Discussion

**3.1. The Rotational Behavior with a Fixed Electric Field along the −y Direction.** We begin the study with the simplest case: the rotor is affected by a constant external electric field, as shown in Figure 1a, defined as the 0° state. Because of the molecule's intrinsic dipole, it is expected to rotate counterclockwise. We then analyze the rotation quantitatively by performing torque analyses. During the counterclockwise rotation, the torques caused by the electric field on different atoms in the rotor, as well as the total torque, can be projected onto the rotation axis and calculated using DFT combined with Newtonian mechanics, as shown in our earlier work.[38]

As the interpolation processes the total torque, a more precise total torque profile with a 1° grid of points can be obtained, as shown in Figure 2, confirming more precise rotational velocities and periods. The fitted total torque is zero at 89°, 251°, 271°, and 288°, which is antisymmetric about 89° in the interval of [0°, 180°] and antisymmetric about 271° in the interval of [180°, 360°]. Furthermore, the fitted total torque with an amplitude of 69.84 pN nm has a maximum (minimum) around 15° (165°) instead of 0° (180°) because the electric field has redistributed the charges. The asymmetry at about 0° and 180° of the torque is due to the inequivalence of reversing the electric field: for example, an electric field pointing along the N−N axis and toward the −NH$_2$ group does not simply produce the opposite charging effect of the reversed field pointing toward the −CN group.

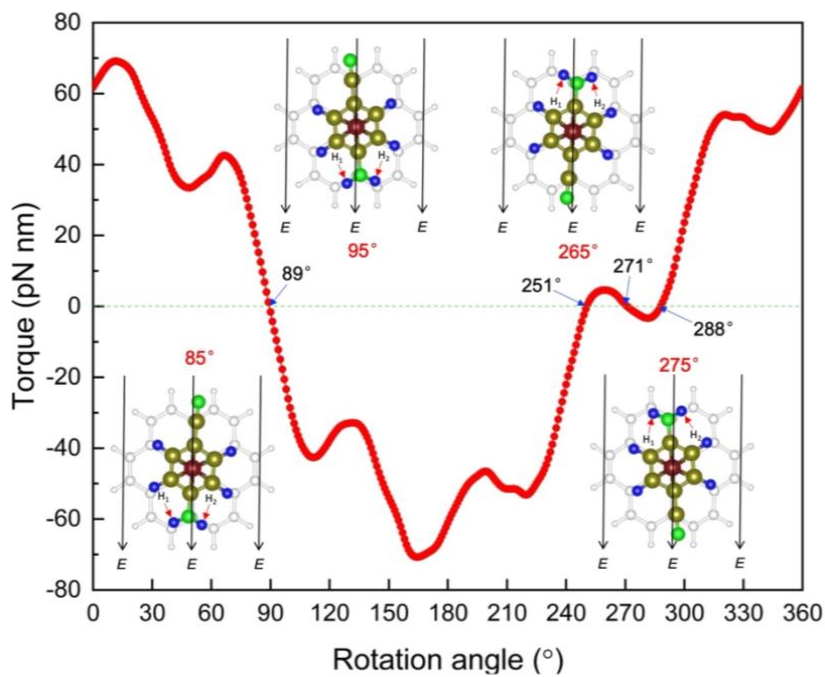

**Figure 2.** Total torque profile with a 1° grid of points processed by interpolation, for a fixed electric field of 0.5 V/Å applied along the −y direction, as a function

of the rotation angle. The inset molecular structures show the system at 85°, 95°, 265°, and 275°, respectively. The blue arrows indicate the orientations with zero torque.

Suppose we release the rotor at 0° with zero initial angular velocity. In that case, it will accelerate counterclockwise, pass 90°, then decelerate to zero angular velocity around 180°, similar to a pendulum, as shown in Figure 3a. Using our formulas, the time spent in each 1° interval is estimated, so the process from 0° to 180° takes 1.48 ps. If the electric field remains unchanged, the rotor will reverse direction clockwise back to its starting point at 0°, as shown in Figure 3b, and continue oscillating back and forth with a period of 2.96 ps, assuming an ideal case without energy loss.

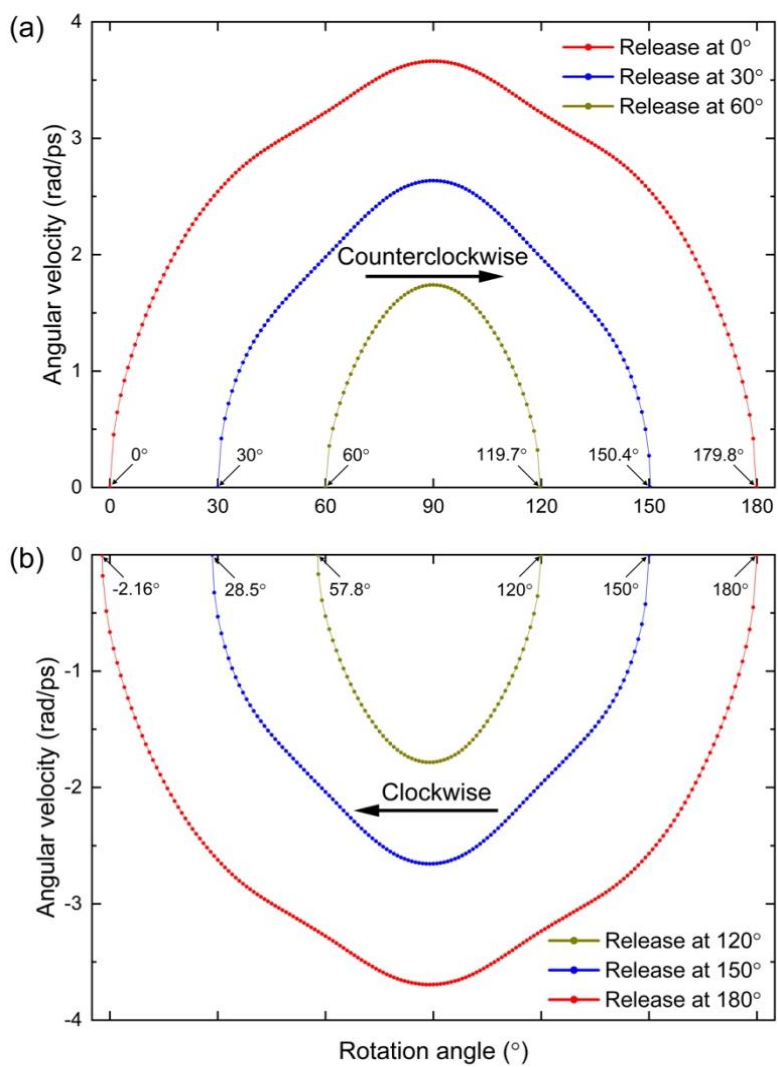

**Figure 3.** Angular velocities of the rotor as a function of rotation angle in a fixed electric field of 0.5 V/Å applied along the −y direction. (a) The rotor rotates counterclockwise during the first half of the rotation cycle, as indicated by the arrow. The red, blue, and dark yellow dotted lines indicate that the rotor is released at 0°, 30°, and 60°, respectively, and stops at 179.8°, 150.4°, and 119.7°. (b) In the second half of the rotation cycle, the rotor rotates clockwise, as indicated by the arrow. The red, blue, and dark yellow dotted lines indicate that the rotor is released at 180°, 150°, and 120°, respectively, and stops at −2.16°, 28.5°, and 57.8°.

The general behavior of dipole alignment in an electric field is well understood. Meanwhile, the proposed dipolar molecular rotor shows that the total energy and torque depend on more than just the net dipole moment. To clarify this difference, a simple rigid dipole in a fixed and constant electric field of 0.5 V/Å has been proposed for comparison, as shown in Figure S1. The total energy of the simple rigid dipole, depending on rotation angle, forms a sine curve with an amplitude of 0.689 eV, but shifted by half a period relative to the standard sine curve, as shown in Figure S2. In contrast, the DFT-calculated total energy of the molecular system, as shown in Figure 9(a) of the earlier work,[38] shows many deviations from the simple rigid dipole. The torque of the simple rigid dipole produces a standard cosine curve with an amplitude of 110.464 pN nm, as shown in Figure S3. However, the DFT-calculated torque is highly asymmetric, as shown in Figure 2. The rotor's response to an external electric field is strongly influenced by charge distribution, anchoring geometry, and structural asymmetry, leading to deviations from the classical rigid-body approximation. These findings highlight the necessity for atomistic-level modeling to accurately capture the complex interactions between molecular structure and field-induced dynamics.

**3.2. The Rotational Behavior with an Alternating Rectangular Electric Field.** To prevent pendulum-like oscillations and keep the rotor rotating counterclockwise continuously, we can reverse the electric field's direction once the rotor reaches 180° and then reverse it again at 0°. This process should be repeated every 2.96 ps, requiring an alternating rectangular electric field (AREF) synchronized with the rotor, as shown in Figure 4. The energy profile of [180°, 360°] will repeat the shape of [0°, 180°], and the same applies to the torque profile. However, this method causes sudden discontinuities as the torque is abruptly reversed. Additionally, some extra energy must be supplied to the rotor to overcome the energy barrier at 180° and to compensate for any energy loss. The rotor will then rotate counterclockwise with a period of 2.96 ps (or less if more energy is supplied). The driving frequency for this scheme is 0.338 THz, which is below the typical electronic excitation thresholds for molecules of this size.[47]

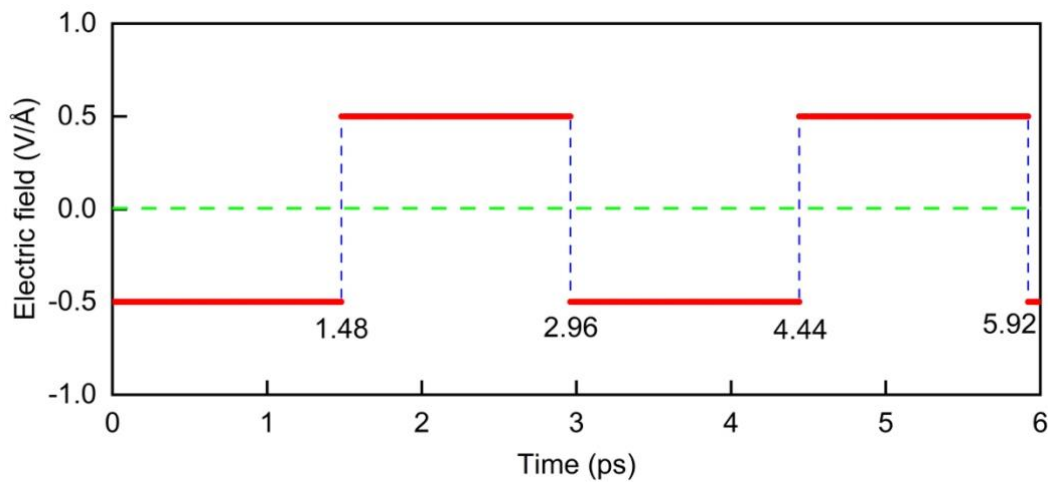

**Figure 4.** Diagrammatic sketch of the alternating rectangular electric field. The positive and negative values indicate the electric fields applied along the +y and −y directions, respectively.

**3.3. The Rotational Behavior with a Cosine Wave Alternating Electric Field Depending on Time.** Another applicable method is to use a cosine wave alternating electric field (AEF) to drive the dipolar molecular rotor. For example, the strength of the electric field can be considered as $E(t) = -\frac{\sqrt{2}}{2}\cos(\frac{2\pi}{T} \times t + \varphi)$.

To compare with the former case and simplify the computational process, we set the period $T$ to 3 ps (instead of the exact value of the alternating rectangular electric field, 2.96 ps) and the initial phase $\varphi$ to 0°. The negative sign indicates that the initial electric field is applied along the −y direction, as shown in Figure 5a inset. Suppose the rotor is released at 0°, the torque $M_Z^0$ acting on the rotor under the instantaneous electric field $E_0$ (where $E_0 = -\frac{\sqrt{2}}{2}\frac{V}{Å} = -0.707$ V/Å ) can be calculated. The time $t_5$ for the rotor to rotate from 0° to 5°, as well as the final angular velocity, can be derived from formulas 4 and 3, respectively. Subsequently, the instantaneous electric field $E_5$ at 5° can be calculated based on the cosine function: $E(t) = -\frac{\sqrt{2}}{2}\cos(\frac{2\pi}{3} \times t_5)$. Similarly, the torque at 5° and the final angular velocity at 10° can also be obtained, extending to the torques and angular velocities at other orientations.

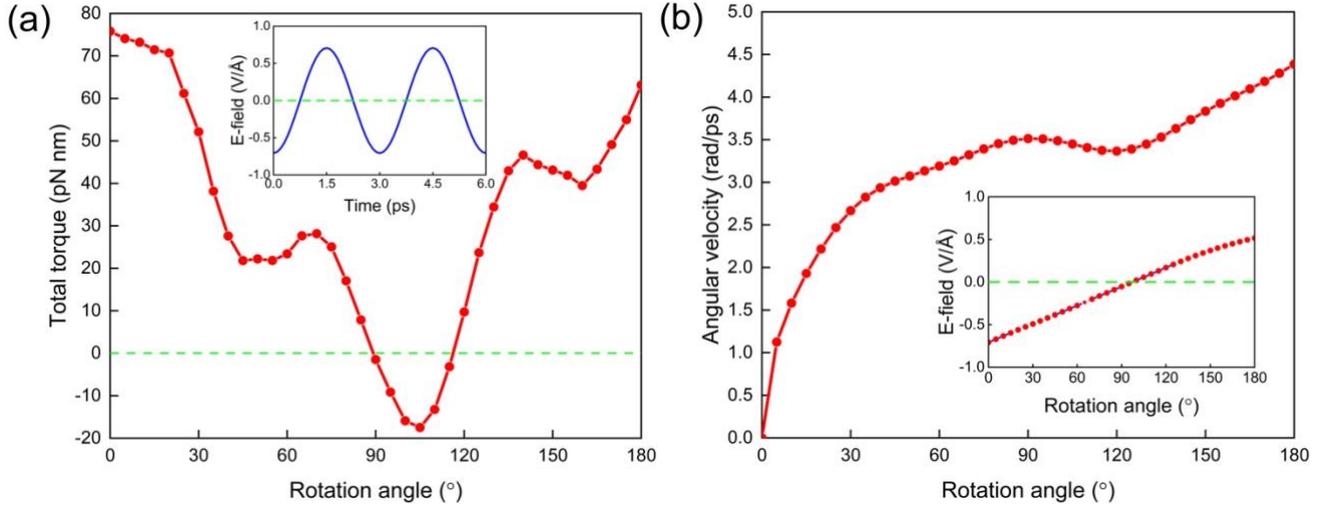

**Figure 5.** (a) The torque profile of the rotor in a cosine wave alternating electric field with an amplitude of $\frac{\sqrt{2}}{2}$ V/Å. We use 5° in each simulation step. The inset shows the electric field profile. The positive and negative values indicate the field applied along the +y and −y directions, respectively. (b) The angular velocity depends on the rotation angle. The inset shows the electric field, which depends on the rotation angle.

Figure 5a shows the torque on the rotor depends on the rotation angle. Accordingly, Figure 5b shows the behavior of the angular velocity, which accelerates in the interval of [0°, 90°] but slightly decelerates in [90°, 120°] as the torque becomes negative during this interval. The estimated total time from 0° to 180° is 1.14 ps, which is less than that in the alternating rectangular electric field. However, the cosine wave electric field with an initial phase of 90° (a sine wave) does not perform well: initially, the field cannot drive the rotor because the initial torque is zero. As time progresses and the electric field changes to a nonzero value along the y direction, the rotor can be driven, but it begins to rotate clockwise. Afterward, the rotor enters an irregular state involving rotation and oscillation. The results indicate that if we release the rotor at 0° with zero initial angular velocity, the 3 ps period cosine wave AEF with an initial phase of 90° cannot drive the rotor regularly. This is fundamental because the time-dependent electric field requires a specific preset period (3 ps in this case) and is thus difficult to control precisely or use generally in applications. Therefore, we conclude that it is more practical to adopt an electric field that depends on the rotor's rotation angle.

**3.4. The Rotational Behavior with a Cosine Wave Alternating Electric Field Depends on the**

**Rotation Angle.** As mentioned, the time-dependent electric field must be strictly confined to a specific pattern (with a particular period) to drive the rotor to rotate regularly. This is often unfeasible. Therefore, a cosine wave electric field depending on the rotation angle to drive the rotor is proposed, expressed as, e.g., $E(\theta) = -\frac{\sqrt{2}}{2}\cos\theta$. In this case, the total energy of the molecular system exhibits two maxima at 105° and 285°, and two minima at 5° and 185°, as shown in Figure 6. The total energy profile is translationally symmetric due to the characteristic of the applied field, which depends on the rotation angle. The potential energy in the cosine wave field exhibits a negative cosine curve, but its period is half that of the standard cosine curve. The atomic torques projected onto the rotation axis have also been calculated, as shown in Figure 7. Similar to the case under a constant electric field, the torques acting on the N atom of the −CN group and the H atoms of the −NH$_2$ group contribute the most. The total torque remains mostly positive during the rotation process, except for two small intervals [90°, 113°] and [271°, 293°], indicating that the rotor can rotate counterclockwise under the cosine wave electric field depending on the rotation angle.

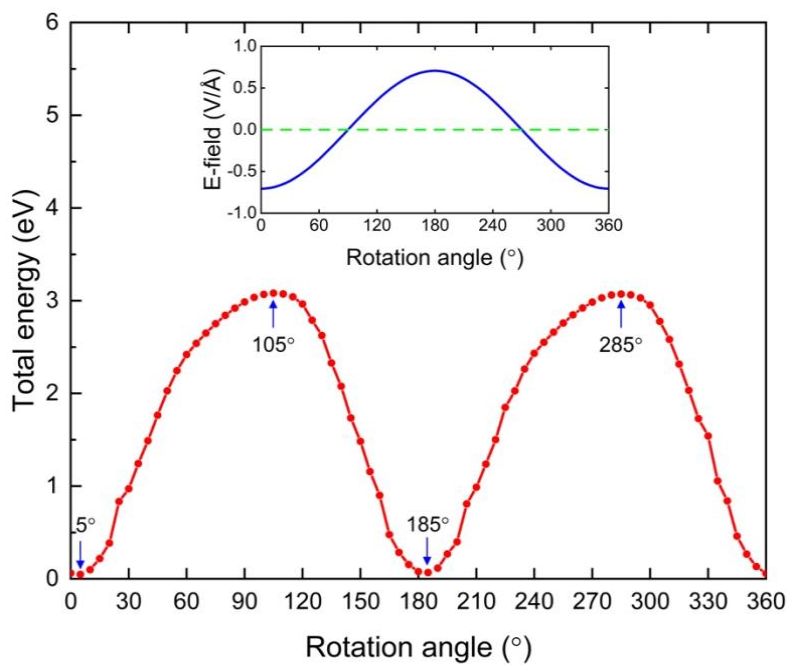

**Figure 6.** Total energy profile of the entire system under a cosine wave electric field, plotted as a function of the rotation angle. The inset shows the cosine wave electric field profile with an amplitude of $\frac{\sqrt{2}}{2}$ V/Å, which depends on the rotation angle. The positive and negative values indicate the field applied along the +y and −y directions, respectively. The minimum of the total energy profile is set to zero.

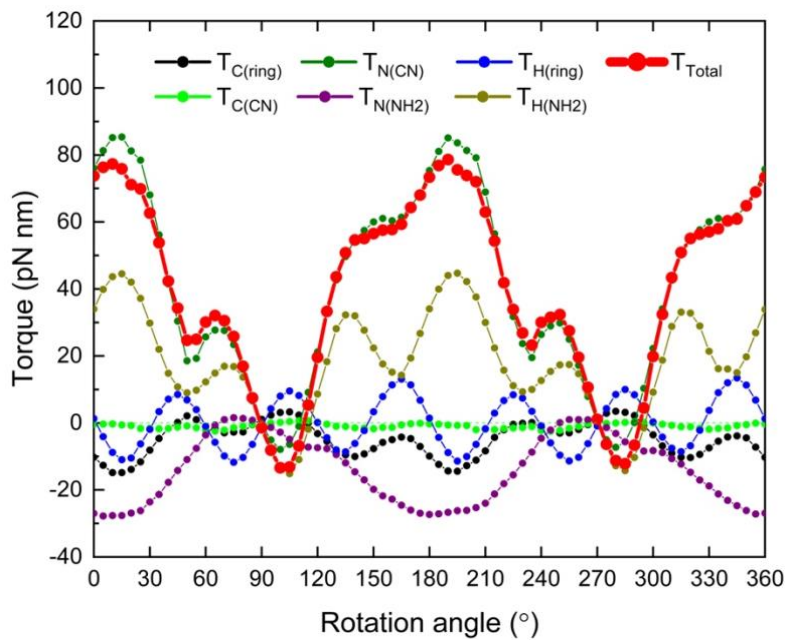

**Figure 7.** Total torque and atomic torques projected onto the rotation axis as functions of the rotation angle. The rotor is driven by a cosine wave electric field that depends on the rotation angle. The black, green, olive, purple, blue, dark yellow, and bold red dotted lines represent the torques on the C atoms of the ring, the C atom of the −CN group, the N atom of the −CN group, the N atom of the −NH$_2$ group, the H atoms attached to the edge of the ring, the H atoms of the −NH$_2$ group, and all atoms of the rotor, respectively.

The total torque profile with a 1° grid of points is also available, as shown in Figure S4. As shown in Figure 8, suppose released at 0° with the applied field shown in the inset of Figure 6, the rotor takes 1.626 ps to rotate from 0° to 360°, reaching a maximum angular velocity of 6.9 rad/ps at 360°. Similarly, if the rotor is released at 30°, 60°, 120°, and 150°, it can also rotate counterclockwise, with final velocities of 6.89, 6.89, 6.90, and 6.90 rad/ps at 30°, 60°, 120°, and 150° of the next cycle, respectively. The total times are 1.92, 2.27, 1.80, and 1.60 ps, respectively. All the curves of the angular velocity profile show two local maxima at 90° and 271°, and two minima at 114° and 294°. If resistance, such as from a car accelerating or climbing a slope, were present, it would produce usable power, preventing infinite acceleration. In that case, the cycle period would be longer than 1.60 ps, and the final angular velocity would likely be lower than 6.9 rad/ps at 360°.

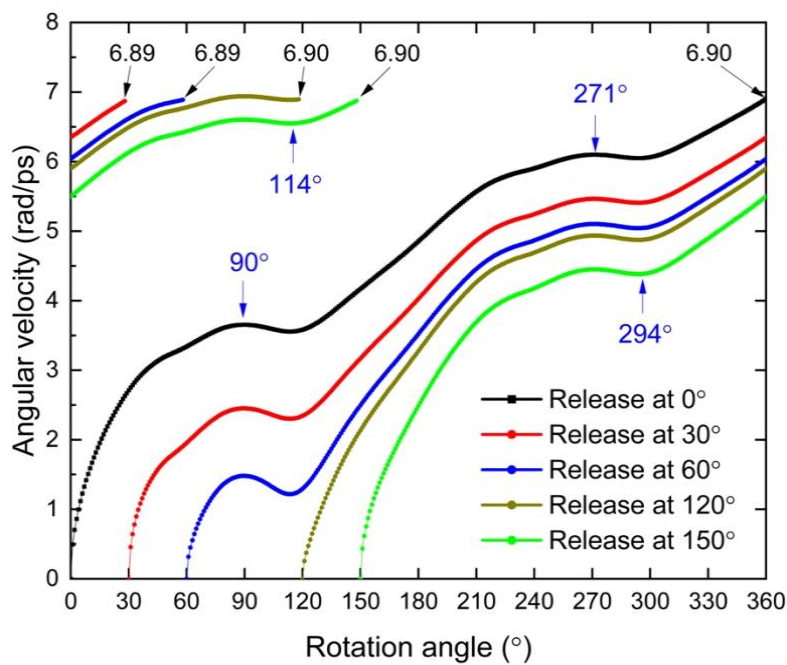

**Figure 8.** Angular velocities of the rotor as a function of the rotation angle for a cosine wave electric field depending on the rotation angle. The black, red, blue, dark yellow, and green dotted lines show the rotor released at 0°, 30°, 60°, 120°, and 150°, respectively. The blue arrows represent the local maxima and minima, and the black arrows represent the final velocities of each cycle.

The angle-dependent electric field can activate the rotor and produce extra power, indicating that the rotor can function as a nanoscale motor. Once it begins to move, the rotating rotor can transfer mechanical energy to nearby components, similar to how a motor turns wheels, providing a potential source of continuous power at the nanoscale. The proposed angle-dependent electric field is a theoretical concept designed to explore the maximum controllability of the rotor. Although this ideal control scheme has not yet been realized experimentally, it serves as a conceptual benchmark for assessing the feasibility of electrostatic actuation at the molecular level. It emphasizes the potential of using angle-dependent field modulation for directional control and sustained rotation. This idea could guide future advancements in feedback-based control systems or nanoelectrode-based field setups, enabling precise manipulation of molecular rotors for nanoscale mechanical tasks.

**3.5. MLMD Simulations and Experimental Outlook.** The simulations above combine electrostatic forces from static DFT calculations with a classical rigid-body approximation to model how molecules rotate under various electric fields. This method effectively captures the main torque contributions at the atomic level; however, it overlooks other effects, such as time-dependent electron responses, nonadiabatic effects,[48] and electronic excitation.[49] As a result, some essential physical behaviors may be missed, especially under strong fields, rapid switching, or near-resonant frequencies, where electronic effects become dominant. To address these limitations, additional MLMD simulations were performed, providing a more realistic view of the rotor's behavior beyond the rigid-body approximation. These MLMD simulations applied a fixed electric field of 0.5 V/Å along the −y direction for 10 ps, at temperatures of 0, 50, 100, and 300 K, respectively. The variations of the total energy over time from these simulations at different temperatures are shown in Figures S5−8. As illustrated in the simulation animations, the rotor remains intact without reconstruction or damage over the 10 ps duration, indicating that the rotor is thermodynamically stable under the applied electric field across all tested temperatures, as shown in Movies S1−S4.

The previously reported caltrop-like rotor with a large dipole moment can reach an equilibrium configuration using DFTB-D molecular dynamics simulations performed at room temperature.[38] The sandwich rotor investigated here has a much smaller dipole moment; the torque analysis predicts a full 180° rotation in about 1.48 ps under an electric field of 0.5 V/Å applied along the −y direction, while none of the MLMD simulations at 0, 50, 100, or 300 K showed the rotor complete a half-cycle rotation (0° to 180°) within 10 ps. This occurs because the torque from the applied electric field is not strong enough to overcome the rotational energy barrier. The rotor might achieve partial or full rotation under a stronger electric field or with a longer simulation time.

The differences between the DFT-based torque analysis and the MLMD trajectories highlight the important roles of rotational energy barriers, stochastic fluctuations, and friction effects in rotational dynamics. As shown in Figures S9 and S10, the total energy and torque profiles computed with the MLMD force field show clear deviations from both the static DFT calculations and the simple rigid-dipole model. This comparison suggests that stochastic fluctuations and friction-like effects play a more dominant role than the rotational energy barrier in determining the observed motion, as thermal energy alone is insufficient to overcome the potential barrier without these additional influences. These results indicate that rotors with larger dipole moments or those exposed to stronger fields are more likely to undergo significant angular displacement within practical time scales. The current MLMD approach is limited to constant electric fields because of the fixed nature of the force field. Modeling rotation under time-dependent or angle-dependent fields would require extensive retraining of the machine learning potential or the use of real-time TDDFT methods, which are beyond the scope of this work.

Direct observation of picosecond-scale molecular rotation remains experimentally challenging. Although scanning tunneling microscopy (STM) has been successfully used to image and manipulate individual rotating molecules,[13-17,19,36] its limited temporal resolution makes it hard to resolve ultrafast rotational dynamics. Nonetheless, STM remains a valuable tool for investigating rotor dynamics in systems with low-frequency motion under weak electric fields. Additionally, inelastic electron tunneling spectroscopy (IETS) provides a promising technique for detecting electronic excitations related to molecular rotation.[16] Although these experimental approaches are technically challenging, they provide feasible ways to explore the dynamic behavior predicted by our simulations. We expect these insights will serve as a foundation for guiding future experimental efforts in nanoscale molecular rotor research.

## 4. Conclusions

A dipolar molecular rotor built on graphene has been studied using density functional theory combined with Newtonian mechanics and MLMD simulations. We demonstrated that the dipolar rotor can be driven by external electric fields. The rotor can rotate around the rotation axis through the Cr atom, whose torques and angular velocities have been calculated. The torque profile has symmetric structures around 90° and 270° points, respectively. In a fixed electric field of 0.5 V/Å applied along the −y direction, the rotor can start from 0° with zero angular velocity, accelerate counterclockwise past 90°, then decelerate near 180° to zero angular velocity in 1.48 ps. It then reverses clockwise back to 0°, continuing oscillations with a period of 2.96 ps. This shows the rotor can rotate periodically in an alternating rectangular electric field with that period. Doubling the electric field strength, while keeping the shape constant, nearly doubles the torque and the angular acceleration (since $M_z = I_{zz} * \alpha$), resulting in faster rotation and quicker slowdown. The time $t$ needed to reach a specific angle $\theta$ from rest can be expressed as $t = \sqrt{2I_{zz}\,\theta/M_z}$, which is reduced by a factor of $1/\sqrt{2}$ when the torque doubles. Assuming a proportional relationship between torque $M_z$ and electric field strength, the time scales as $t = cont/\sqrt{E}$. Since this is uniformly accelerated motion with each 1° step with finite initial velocity, the period will be shorter than the initial period $1/\sqrt{2}$ when the torque doubles. Lower electric field strength would require longer periods. This implies that the electric field's overall strength influences the period, enabling continuous, unidirectional rotation. Conversely, higher electric field strengths make convergence in DFT calculation more challenging. A cosine wave AEF with the same effective strength and period, when starting at an initial phase of 0°, induces a rapid and continuous counterclockwise rotation of the molecular rotor. In contrast, a cosine wave AEF beginning at a 90° phase fails to produce a regular rotational motion. Interestingly, if the rotor is driven by a cosine wave electric field depends on its rotation angle with an initial phase of 180°, the total torque is nearly always positive in the counterclockwise direction, allowing continuous rotation and potential power output. Releasing the rotor at different starting points can yield a terminal velocity of approximately 6.90 rad/ps after one cycle. MLMD simulations indicate that the rotor remains thermodynamically stable under the applied electric field across a broad temperature range.

Although a model that does not include the environment has been proposed in this work, it provides a possible path to understand the rotation mechanism of the dipolar molecular rotor at the nanoscale induced by the electric field, and it also provides a reference for designing a useful motor for delivering work in real-world applications.

**Supporting Information**

> Movie S1. Top view animation of the molecule for the MLMD simulations from the 0° state in a fixed and constant electric field of 0.5 V/Å along the −y direction at 0 K; the interval between frames is 0.02 ps (MP4).
> Movie S2. Side view animation of the molecule for the MLMD simulations from the 0° state in a fixed and constant electric field of 0.5 V/Å along the −y direction at 0 K; the interval between frames is 0.02 ps (MP4).
> Movie S3. The same as Movie S1 but at 50K (MP4).
> Movie S4. The same as Movie S2 but at 50K (MP4).
> Movie S5. The same as Movie S1 but at 100K (MP4).
> Movie S6. The same as Movie S2 but at 100K (MP4).
> Movie S7. The same as Movie S1 but at 300K (MP4).
> Movie S8. The same as Movie S2 but at 300K (MP4).

I.   One of the Input Files at Equilibrium State

   POSCAR of the rotor at 0° state configuration

## II. Simple Rigid Dipole Model

Figure S1. A simple rigid dipole in a fixed and constant applied electric field; the green and blue spheres represent the positive and negative point charges, respectively, the red sphere indicates the pivot point, the black arrow indicates the dipole moment $\vec{p}$, $\rho$ indicates the angle between $\vec{p}$ and the electric field $\vec{E}$, and $\theta$ indicates the rotation angle.

Figure S2. Total energy profile of the simple rigid dipole in the fixed field of 0.5 V/Å.

Figure S3. Torque profile of the simple rigid dipole in the fixed field of 0.5 V/Å.

## III. Torque Profile

Figure S4. Total torque profile with a 1° grid of points processed by interpolation, for a cosine wave applied electric field depending on the rotation angle with an initial phase of 0°, as a function of the molecular rotation angle, the blue arrows indicate the orientations with zero torque.

## IV. MLMD Simulations

Figure S5. Time-dependent total energy profile of the molecule obtained from MLMD simulations at 0 K.

Figure S6. Time-dependent total energy profile of the molecule obtained from MLMD simulations at 50 K.

Figure S7. Time-dependent total energy profile of the molecule obtained from MLMD simulations at 100 K.

Figure S8. Time-dependent total energy profile of the molecule obtained from MLMD simulations at 300 K.

Figure S9. Time-dependent total energy profile of the molecule obtained from MLMD simulations at 0 K, sampled at 0.02 ps intervals, the minimum of the total energy profile is set to zero, it exhibits pronounced deviations from both the static calculations and the simple rigid-dipole model.

Figure S10. Time-dependent torque profile of the rotor obtained from MLMD simulations at 0 K, sampled at 0.02 ps intervals, the rotor reaches its maximum rotation angle (~45°) at about 1.2 ps, then continues with sustained oscillations, positive torque accelerates the rotor, while negative torque

decelerates it, it also exhibits pronounced deviations from both the static calculations and the simple rigid-dipole model.

## Acknowledgments

We thank Rui-Qin Zhang, Dao-Xin Yao, Rundong Zhao, Jiesen Li, and Michel A. Van Hove for their valuable discussions and insightful suggestions. The work was supported by the National Natural Science Foundation of China (Grant Nos. 12304079, 61801520, and 21703190), the program for scientific research start-up funds of Guangdong Ocean University (Grant No. YJR22028), the grant from the University Research & Development Project of Shenzhen Polytechnic University (Grant No. 513-602431Y003P), the Hunan Provincial Natural Science Foundation of China (Grant No. 2024JJ5111), and the Natural Science Foundation of Changsha City (Grant No. kq2208055). We also acknowledge the computing resources provided by the Tianhe2-JK cluster at the Beijing Computational Science Research Center (CSRC) and the Tianhe cluster at the National Supercomputer Center in Guangzhou.

## Author contributions

W.L. performed the calculations and analyzed the results. W.L. also drafted the manuscript, and all authors reviewed and provided comments.

## Additional information

Competing Interests: The authors declare no competing interests.

For Table of Contents use only

8.3cm ×3.5cm

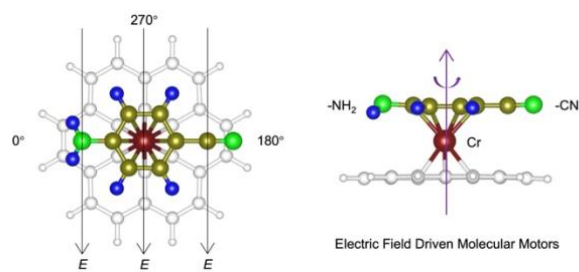

A high-resolution TIF file has been uploaded to the ACS submission system.

Supporting Information for

# A Molecular Rotor Driven by an Electric Field on Graphene

Wanxing Lin,[†, ‡] Xiaobo Li,[§] and Yan-Ling Zhao [*, ‡, ||]

[†] School of Materials Science and Engineering, Guangdong Ocean University, Yangjiang 529500, P. R. China

[‡] Department of Physics, City University of Hong Kong, Hong Kong SAR 999077, P. R. China

[§] School of Physics and Optoelectronic Engineering, Hainan University, Haikou 570228, P. R. China

[||] Advanced Energy Storage Technology Research Center, Shenzhen Polytechnic University, Shenzhen 518055, P. R. China

[*] E-mail: zhaoyanling@szpu.edu.cn.

## CONTENTS

## I. One of the Input Files at Equilibrium State

POSCAR of the rotor at 0° state configuration

VASP @ Device Studio
1.0
25.8241199312   0.0000000000   0.0000000000
0.0000000000   25.8241199312   0.0000000000
0.0000000000   0.0000000000   25.0000000000
Cr C N H
1 31 2 18
Cartesian
12.9130115509 12.9180784225 13.9719200134
11.6322149316 9.3618100450 12.3603908012
10.3865778673 11.4682851884 12.3091088468
9.1830816423 13.5986798926 12.2637370233
14.1254424892 9.3629651730 12.3566039968
12.8781429600 11.4713430882 12.3121387365
11.6323823500 13.6238095958 12.3051461376
10.4302850887 15.7588718930 12.3266482513
15.3658225545 11.4715208233 12.2818014617
14.1206546951 13.6240310841 12.2638209753
12.8781429600 15.7816917405 12.3362094097
16.5699650709 13.5985732009 12.2556649825
15.3238458281 15.7577541923 12.3166307374
10.4312638215 10.0532243255 12.3341118998
9.1834234851 12.2122692566 12.2655931453
12.8781429600 10.0321443465 12.3485586924
11.6323798496 12.1898207592 12.3114019880
10.3859126364 14.3437301500 12.3032488601
15.3241905812 10.0572835290 12.3248702560
14.1206220026 12.1903688284 12.2699861712
12.8781429600 14.3427553733 12.2993942839
11.6311509940 16.4508905729 12.3505885461
16.5699906139 12.2161574845 12.2572225158
15.3657452289 14.3431925276 12.2758835167
14.1250460803 16.4517271671 12.3460230526
12.1469726562 14.1431264877 15.5554504395
14.2888565063 12.9180784225 15.6007165909
11.4242630005 12.9180784225 15.6041402817
13.5652894974 14.1556730270 15.5575761795
12.1472387314 11.6937074661 15.5601367950
13.5652084351 11.6813287735 15.5617103577
15.7149457932 12.9180784225 15.6028327942

16.8862915039 12.9180784225 15.6375675201
10.0282220840 12.9180784225 15.6713495255
11.6015005112 15.0872726440 15.4751329422
11.6023740768 10.7488307953 15.4854068756
14.1116895676 15.0979652405 15.4866619110
14.1114940643 10.7386131287 15.4967174530
11.6287982894 8.2690620085 12.3796072800
8.2342420908 14.1406299507 12.2284266107
14.1312857856 8.2706232472 12.3885457118
9.4861438377 16.3096752504 12.3202991395
17.5186414062 14.1396189244 12.2750311097
16.2695367109 16.3051086178 12.3296277621
9.4874841108 9.5017516920 12.3252032047
8.2351455602 11.6692118683 12.2311700455
16.2699081337 9.5098516894 12.3349130163
11.6264975791 17.5435512474 12.3712891137
17.5187383146 11.6752401027 12.2775112122
14.1300551706 17.5439534900 12.3798610603
9.5973958969 13.7539386749 15.2778759003
9.5971632004 12.0822801590 15.2779636383

## II. Simple Rigid Dipole Model

A simple rigid dipole in a fixed and constant electric field of 0.5 V/Å applied along the −y direction, as shown in Figure S1. The distance between the two point charges is 2r, with the pivot at their midpoint. The total energy is

$$U = -\vec{p}\vec{E} = -pE\cos\rho = -pE\sin\theta$$

where $\vec{p}$ is the dipole moment vector of the rotor, $\vec{E}$ is the applied electric field, $\rho$ is the angle between $\vec{p}$ and $\vec{E}$, and $\theta$ is the rotation angle, going from 0 to 360°; therefore, $\rho = \theta - 90°$. In this work, the absolute dipole moment value of the molecular rotor $\vec{p}$ is 1.379 eÅ (6.623 D).

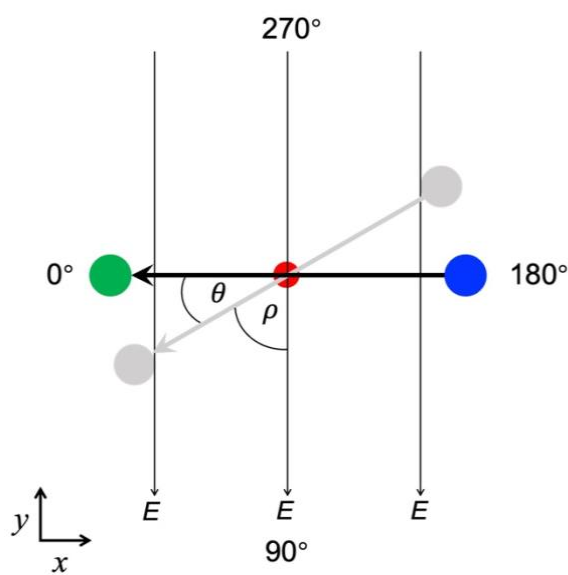

**Figure S1.** A simple rigid dipole in a fixed and constant applied electric field; the green and blue spheres represent the positive and negative point charges, respectively, the red sphere indicates the pivot point, the black arrow indicates the dipole moment $\vec{p}$, $\rho$ indicates the angle between $\vec{p}$ and the electric field $\vec{E}$, and $\theta$ indicates the rotation angle.

To compare the simple rigid dipole and the molecular rotor cases, their molecular moment values and the applied field should be equal. In this scenario, the total energy of the rigid dipole in the field is

$$U = -1.379 \times 0.5 \cos\rho = -0.689 \times \sin\theta \text{ (eV)}.$$

The total energy profile of the simple rigid dipole in the field exhibits a sine curve with an amplitude of 0.689 eV, but shifted by half a period relative to the standard sine curve, as shown in Figure S2.

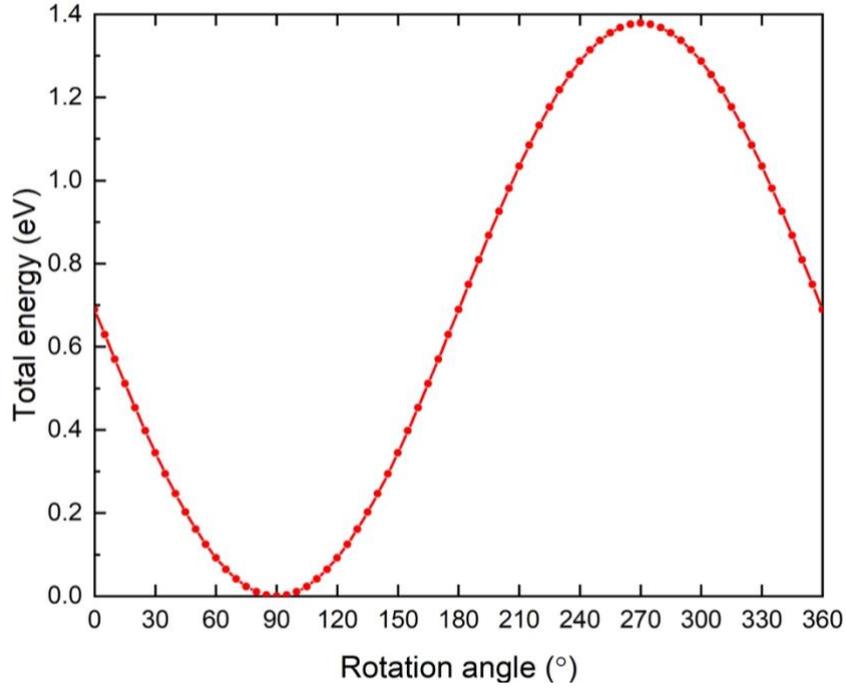

**Figure S2.** Total energy profile of the simple rigid dipole in the fixed field of 0.5 V/Å.

Furthermore, the torque depends on the rotation angle

$$M_z = 2rqE \cos\theta = 0.689 \cos\theta \text{ (eV)},$$

which is a standard cosine curve with an amplitude of 110.464 pN nm (1 eV = 160.218 pN nm), as shown in Figure S3.

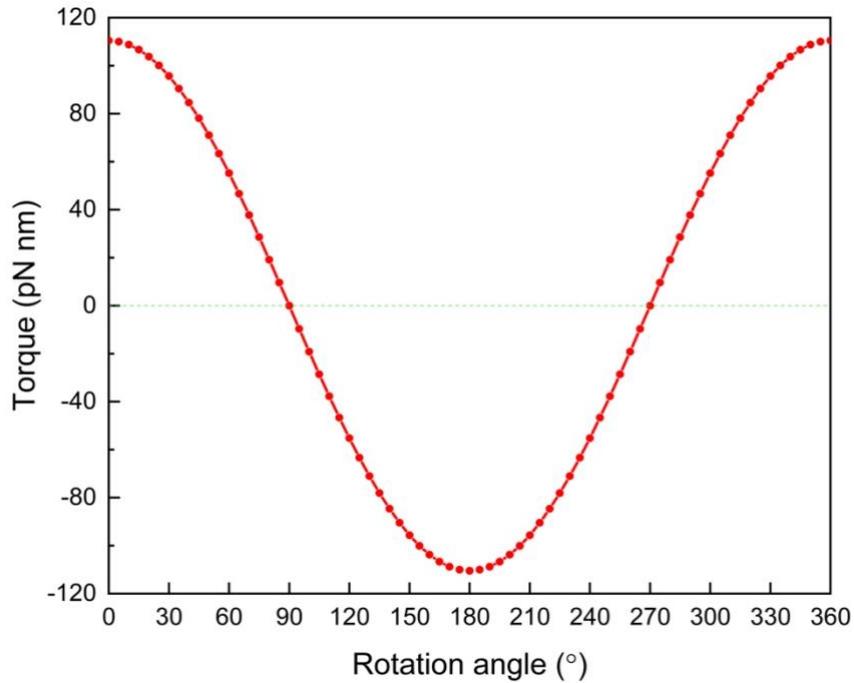

**Figure S3.** Torque profile of the simple rigid dipole in the fixed field of 0.5 V/Å.

III. Torque Profile

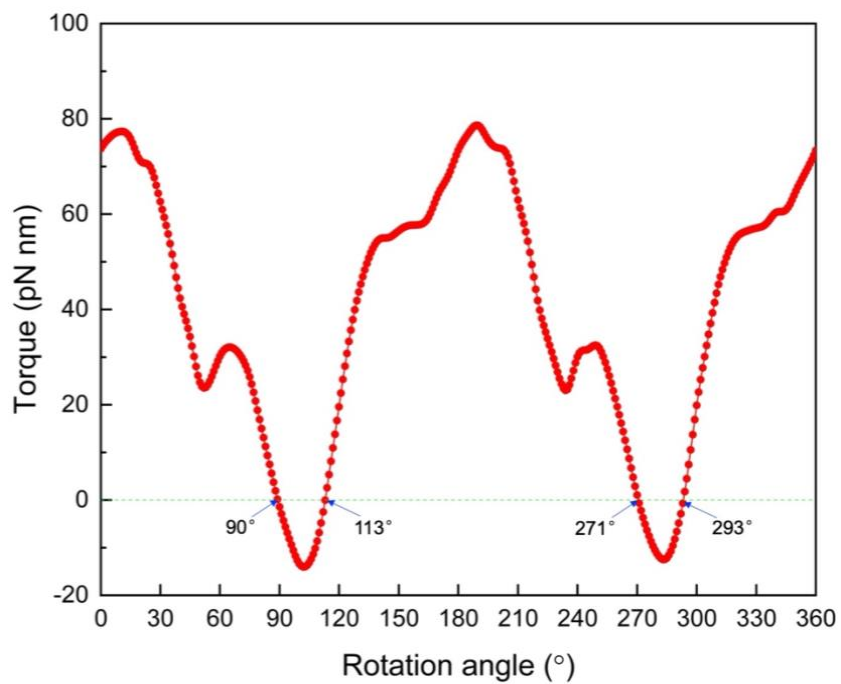

**Figure S4.** Torque profile with a 1° grid of points processed by interpolation, for a cosine wave applied electric field depending on the rotation angle with an initial phase of 0°, as a function of the molecular rotation angle, the blue arrows indicate the orientations with zero torque.

## IV. MLMD Simulations

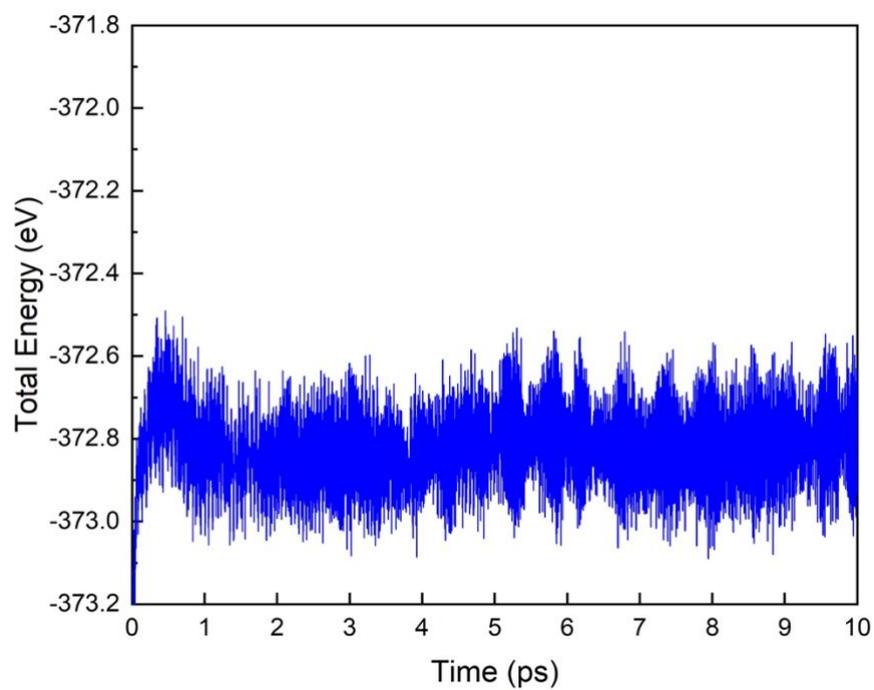

**Figure S5.** Time-dependent total energy profile of the molecule obtained from MLMD simulations at 0 K.

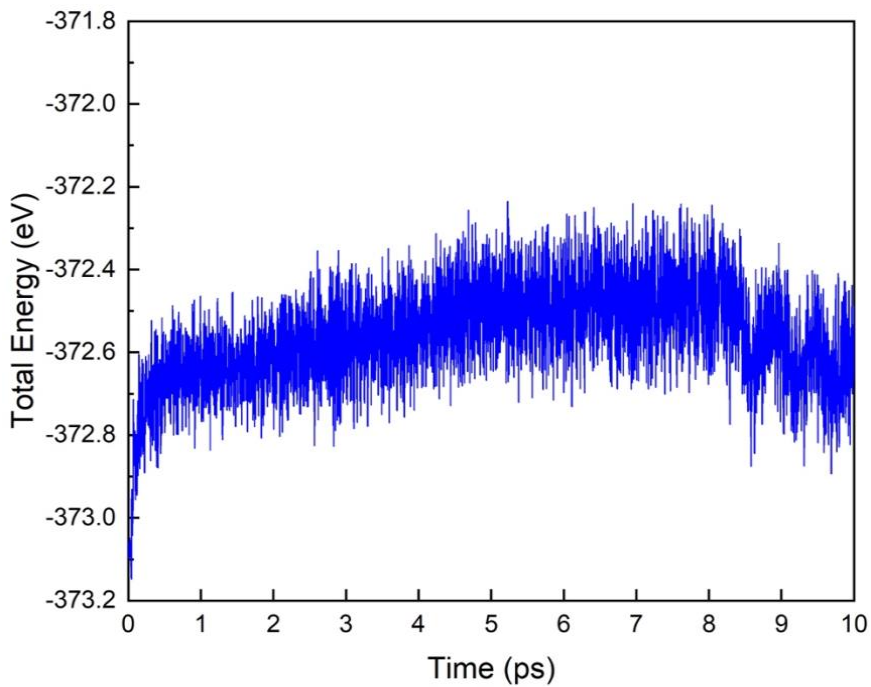

**Figure S6.** Time-dependent total energy profile of the molecule obtained from MLMD simulations at 50 K.

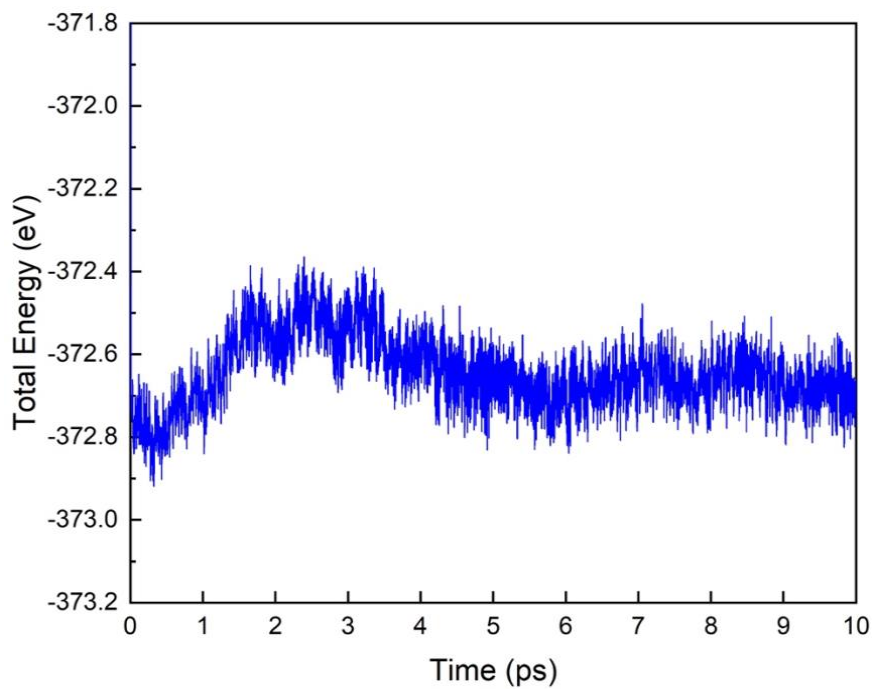

**Figure S7.** Time-dependent total energy profile of the molecule obtained from MLMD simulations at 100 K.

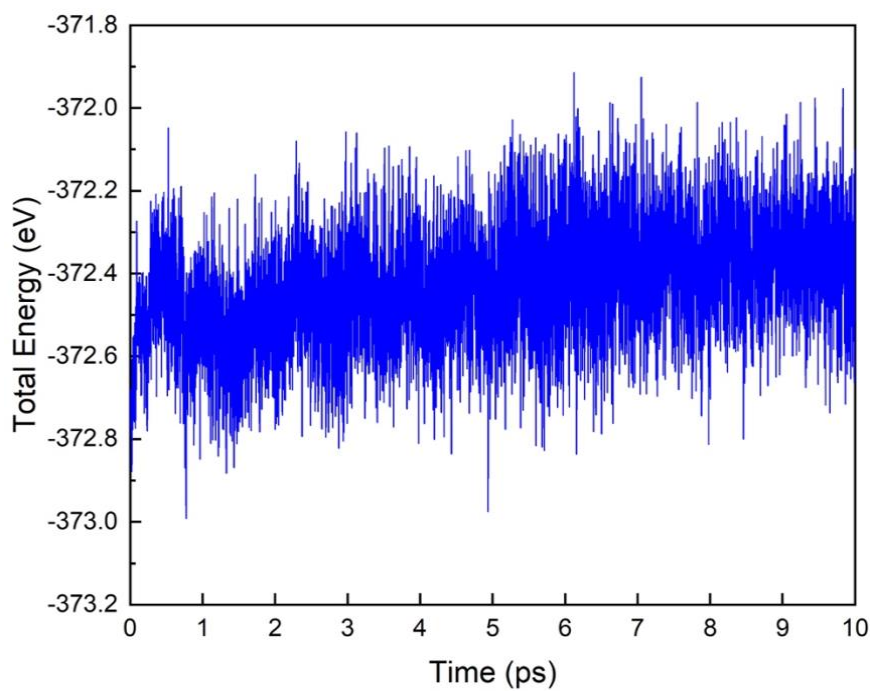

**Figure S8.** Time-dependent total energy profile of the molecule obtained from MLMD simulations at 300 K.

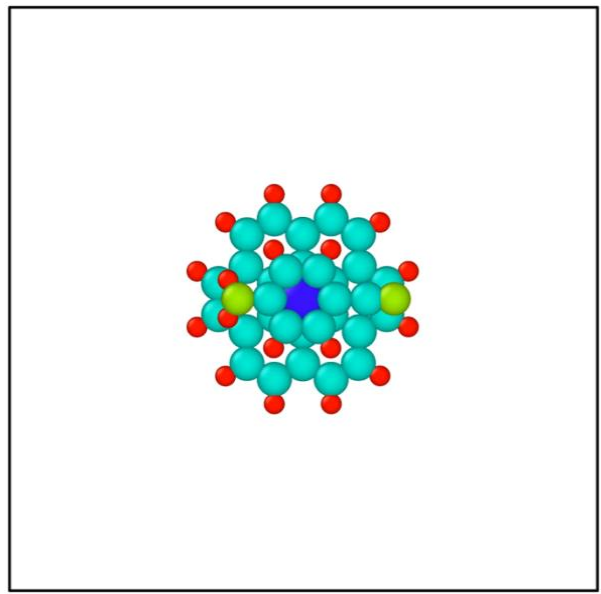

**Movie S1**. Top view animation of the molecule for the MLMD simulations from the 0° state in a fixed and constant electric field of 0.5 V/Å along the −y direction at 0 K; the interval between frames is 0.02 ps (MP4).

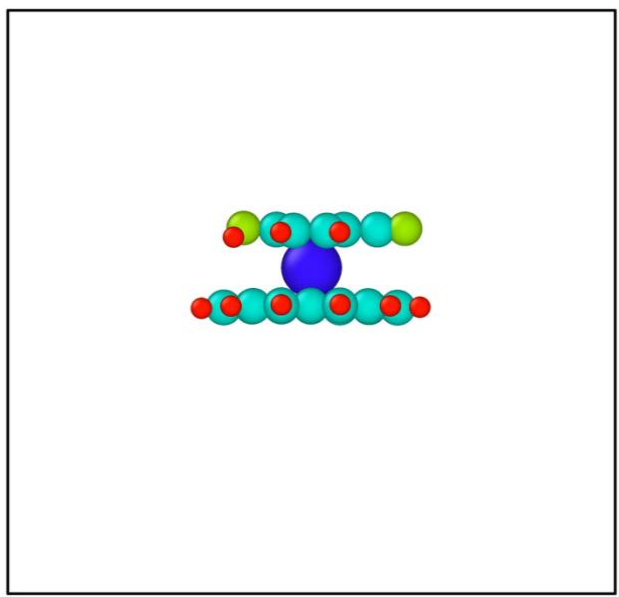

**Movie S2**. Side view animation of the molecule for the MLMD simulations from the 0° state in a fixed and constant electric field of 0.5 V/Å along the −y direction at 0 K; the interval between frames is 0.02 ps (MP4).

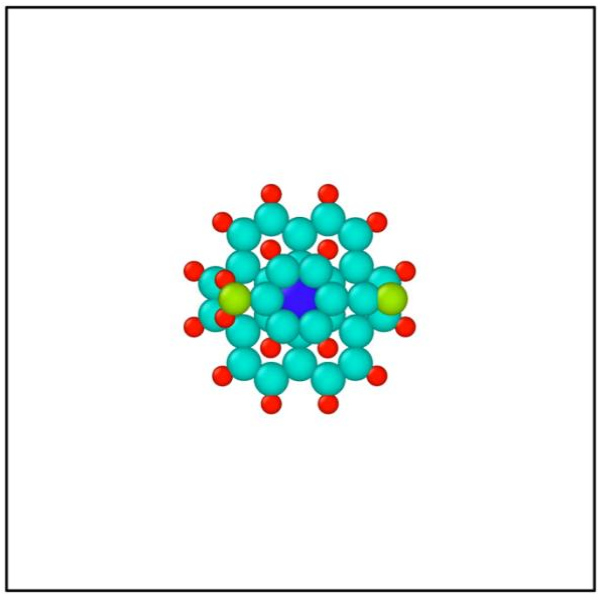

**Movie S3**. The same as Movie S1 but at 50K (MP4).

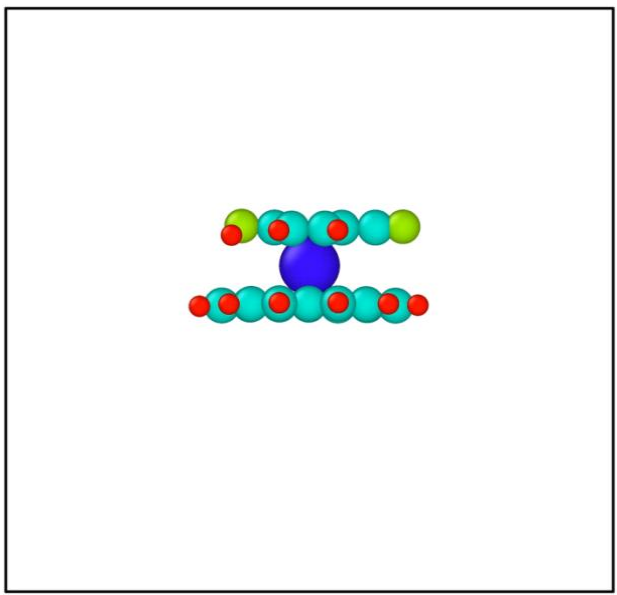

**Movie S4**. The same as Movie S2 but at 50K (MP4).

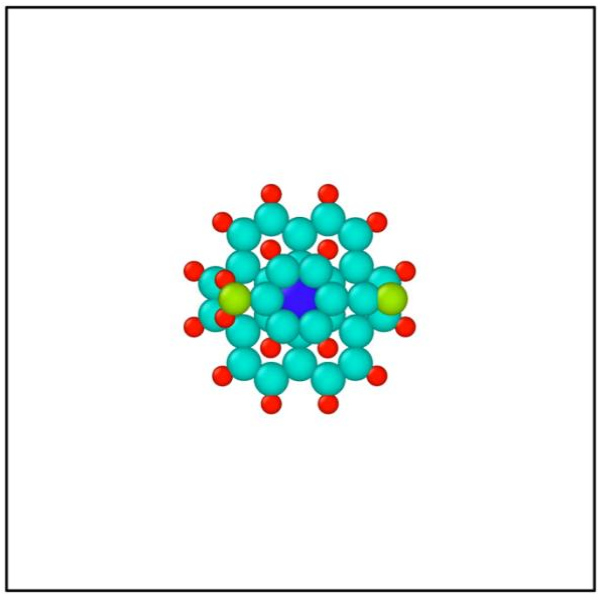

**Movie S5**. The same as Movie S1 but at 100K (MP4).

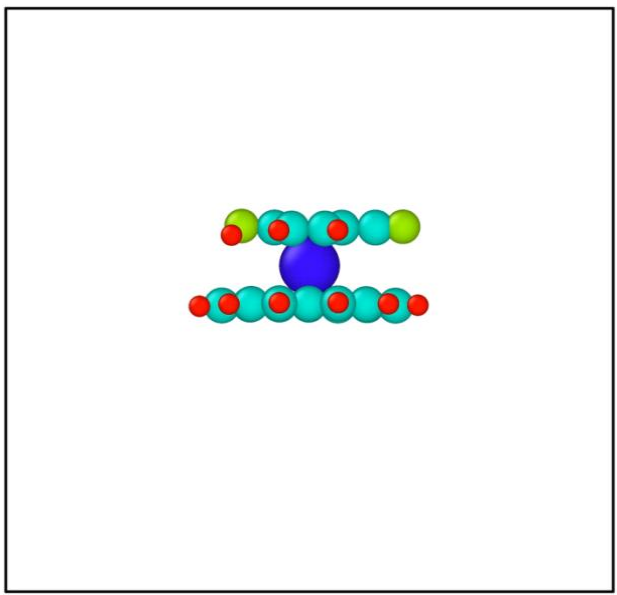

**Movie S6**. The same as Movie S2 but at 100K (MP4).

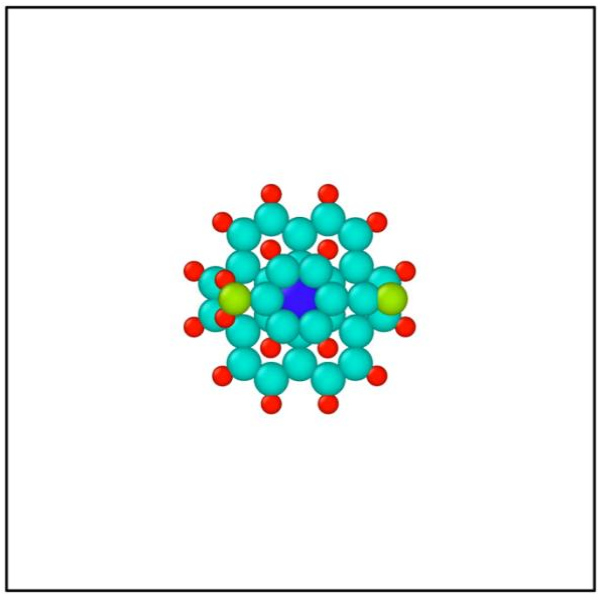

**Movie S7**. The same as Movie S1 but at 300K (MP4).

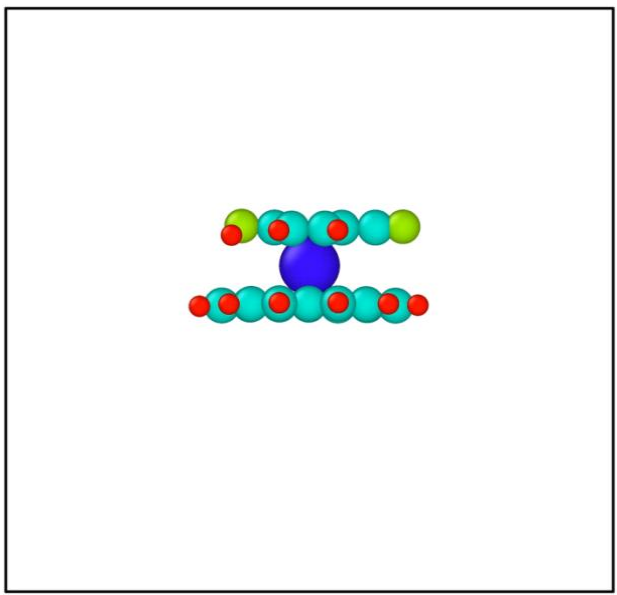

**Movie S8**. The same as Movie S2 but at 300K (MP4).

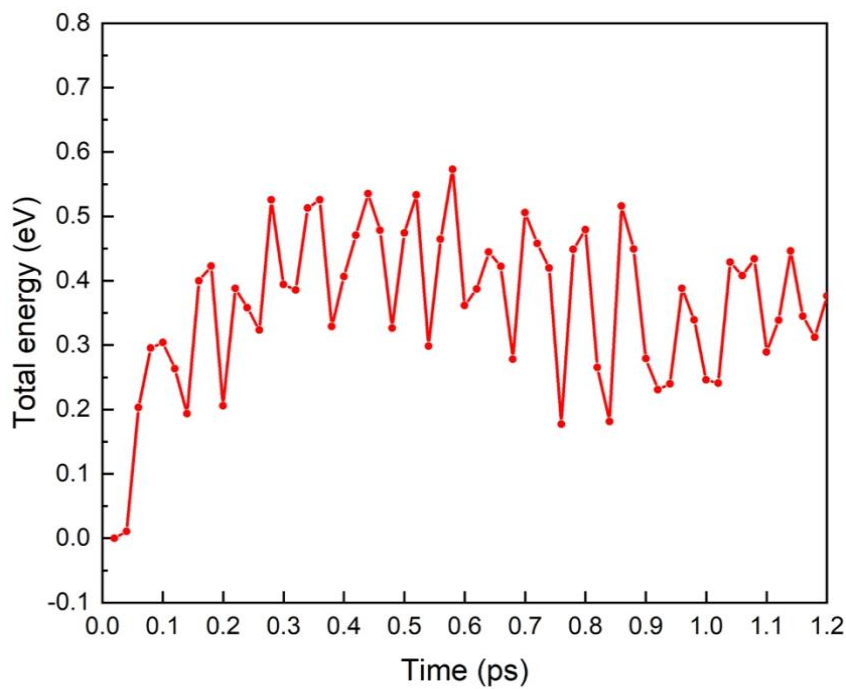

**Figure S9.** Time-dependent total energy profile of the molecule obtained from MLMD simulations at 0 K, sampled at 0.02 ps intervals, the minimum of the total energy profile is set to zero, it exhibits pronounced deviations from both the static calculations and the simple rigid-dipole model.

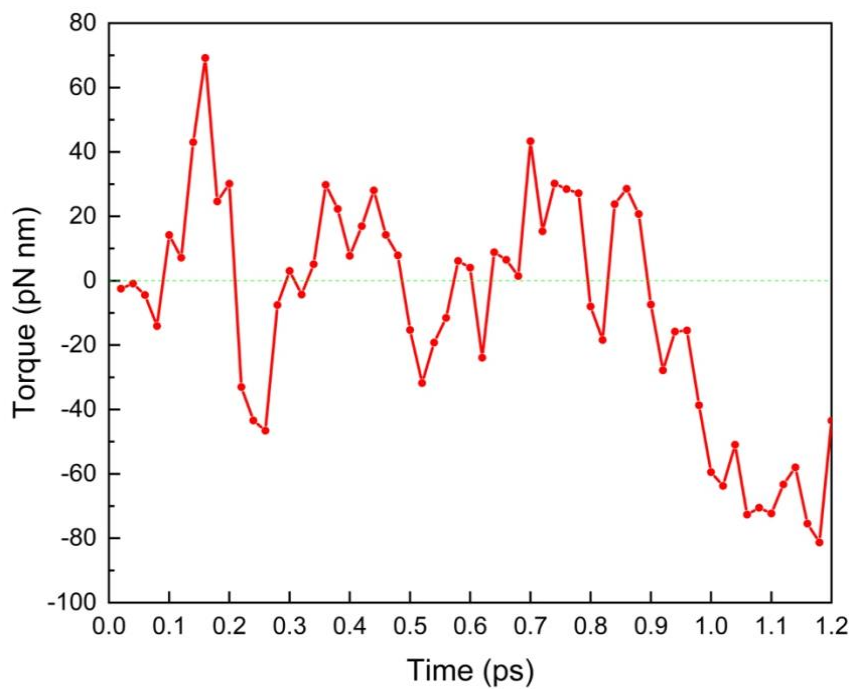

**Figure S10.** Time-dependent torque profile of the rotor obtained from MLMD simulations at 0 K, sampled at 0.02 ps intervals, the rotor reaches its maximum rotation angle (~45°) at about 1.2 ps, then continues with sustained oscillations, positive torque accelerates the rotor, while negative torque decelerates it, it also exhibits pronounced deviations from both the static calculations and the simple rigid-dipole model.